\documentclass[english,twocolumn,superscriptaddress,prb,aps,showpacs]{revtex4}
\usepackage[T1]{fontenc}
\usepackage[latin1]{inputenc}
\usepackage{amsmath}
\usepackage{graphicx}
\usepackage{amssymb}

\makeatletter
\makeatletter
\usepackage{bm}

\usepackage{dcolumn}

\usepackage{bm}

\usepackage{endnotes}

\makeatother

\usepackage{babel}
\makeatother
\begin{document}

\author{Alexey A. Kovalev}

\affiliation{Department of Physics, Texas A\&M University, College Station, TX
77843-4242, USA}

\author{Mario F. Borunda }

\affiliation{Department of Physics, Texas A\&M University, College Station, TX
77843-4242, USA}

\author{T. Jungwirth}

\affiliation{Institute of Physics ASCR, Cukrovarnick\'{a} 10, 162 53 Praha 6,
Czech Republic }

\affiliation{School of Physics and Astronomy, University of Nottingham, Nottingham
NG7 2RD, UK}

\author{L. W. Molenkamp}

\affiliation{Physikalisches Institut(EP 3), Universität Würzburg, Am Hubland,
97074 Würzburg, Germany}

\author{Jairo Sinova}

\affiliation{Department of Physics, Texas A\&M University, College Station, TX
77843-4242, USA}

\title{Aharonov-Casher effect in a two dimensional hole gas with spin-orbit
interaction }

\begin{abstract}
We study the quantum interference effects induced by the Aharonov-Casher
phase in a ring structure in a two-dimensional heavy hole (HH) system
with spin-orbit interaction realizable in narrow asymmetric quantum
wells. The influence of the spin-orbit interaction strength on the
transport is analytically investigated. These analytical results allow
us to explain the interference effects as a signature of the Aharonov-Casher
Berry phases. Unlike the previous studies on the electron two-dimensional
Rashba systems, we find that the frequency of conductance modulations
as a function of the spin-orbit strength is not constant but increases
for larger spin-orbit splittings. In the limit of thin channel rings
(width smaller than Fermi wavelength), we find that the spin-orbit
splitting can be greatly increased due to the quantization in the radial
direction. We also study the influence of magnetic field considering
both limits of small and large Zeeman splittings. 
\end{abstract}

\date{\today{}}


\maketitle
Particles propagating through a coherent nanoscale device acquire
a quantum geometric phase which can have important physical consequences.
This geometric phase, known as Berry phase,\cite{Berry:1984} is acquired
through the adiabatic motion of a quantum particle in the system's
parameter space and can have strong effects on the transport properties
due to self-interference effects of the quasiparticles when moving
in cyclic motion. Its generalization to non-adiabatic motion is known
as the Aharonov-Anandan phase.\cite{Aharonov:apr1987} A classical
example of such geometric phases is the Aharonov-Bohm phase acquired
by a particle going around a loop in the presence of a magnetic flux.
An important corollary to this phase is the Aharonov-Casher (AC) phase
arising from the propagation of an electron in the presence of spin-orbit
coupling.\cite{AHARONOV:1984} This novel effect has attracted strong
interest within the spintronic research community which focuses, among
other things, on spin-dependent control through electrical means.\cite{Wolf:nov2001,Zutic:apr2004}

\begin{figure}[t]
 %\centerline{\includegraphics[width=1.0\columnwidth]{cartoon.eps}}
\centerline{\includegraphics[width=0.9\columnwidth]{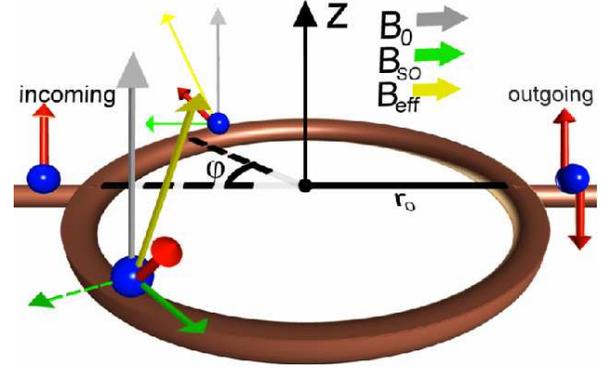}}

\caption{(Color online.) One channel ring of radius $r_{0}$ subject to spin-orbit
coupling in the presence of an additional magnetic field $\mathbf{B_{0}}$.
Electron (hole) spin travelling around the ring acquires phase due
to the applied out-of-plane magnetic field (gray arrow) and the spin-orbit
in-plane magnetic field (momentum dependent, green full-line arrows
for holes and dashed line arrows for electrons) caused by the spin-orbit
interaction. The spin-orbit in-pane magnetic field is different for
holes and electrons.}

\label{RingDraw} 
\end{figure}

Spintronics has made its way into many niche technological applications,
e.g. magnetic memories or MRAM's,\cite{Parkin_2002} using effects
that take place in metals. However, the majority of modern electronic
devices are based on semiconductors and more applications will be
possible when semiconductor devices can employ the spin degree of
freedom as another functional variable in computational processing.
The effects of the AC phase on transport through semiconducting ring-structures
can be tested in two dimensional gas confined to an asymmetric potential
well. Such structures enable an all electrical control of the spins
via the Rashba spin-orbit interaction by changing the gate voltage.\cite{BYCHKOV:1984,ARONOV:jan1993,Choi:aug1997,Nitta:aug1999,Mal'shukov:jul1999}
This spin-interference in a semiconductor ring (see Fig.~\ref{RingDraw})
has been proposed as a way to control spin-polarized currents\cite{Frustaglia:jun2004,Tserkovnyak:2006}
and as a spin-filter.\cite{Molnar:apr2004} Signatures of the Aharonov-Casher
effect have already been experimentally detected,\cite{Morpurgo:feb1998,Konig:feb2006,Bergsten:nov2006}
and more theoretical\cite{Souma:nov2004} and experimental\cite{Habib:2006}
studies have become available recently. %

Spin-interference relies on the spin-splitting and, as a result, the
devices with stronger spin-splitting can provide more control over
the spin. Quantum wells with the spin-orbit interaction proportional
to the cube of the momentum (e.g. with a heavy hole (HH) band)\cite{Zhang:jun2001}
show, in general, larger spin-orbit splittings. We study here the
behavior of the narrow ring in the presence of this cubic spin-orbit
interaction. The analysis of recent experiments \cite{Konig:feb2006,Habib:2006}
shows that the conductance modulations have larger frequency of oscillations
compared to the expected one from a single channel analysis due to
Aharonov-Casher effect and it is within a linear-Rashba multi channel
conductance analysis that agreement is reached.\cite{Konig:feb2006}
In this paper, we analyze whether the larger frequency can be a result
of the cubic spin-orbit interaction in a single channel mode. First,
we develop a theoretical approach based on the assumption of perfect
coupling between leads and the ring. This approach enables us to analytically
calculate the Aharonov-Casher modulations of the conductance as a
function of the spin-orbit splitting. By introducing an external magnetic
field, we also calculate the combined Aharonov-Casher and Aharonov-Bohm
conductance modulations. Finally, we study the influence of the Zeeman
splitting on the conductance.

The 2D Hamiltonian for a single heavy hole (HH) in the presence of
spin-orbit interaction and a magnetic field is given by \begin{equation}
H_{2D}=\frac{1}{2m^{*}}\boldsymbol{\mathbf{\Pi}}^{2}+\frac{1}{2}g\mu_{B}\mathbf{B}\cdot\boldsymbol{\mathbf{\sigma}}+\frac{\alpha}{\hbar}(\sigma_{+}\Pi_{-}^{3}-\sigma_{-}\Pi_{+}^{3})+V(\mathbf{r}),\label{ham}\end{equation}
 where $g$ is the gyromagnetic ratio, $\mu_{B}$ is the Bohr magneton,
$\boldsymbol{\mathbf{\sigma}}$ is the vector of the Pauli spin matrices,
$\boldsymbol{\mathbf{\Pi}}=\mathbf{p}+(e/c)\mathbf{A}$, $\mathbf{\Pi}_{\pm}=\mathbf{\Pi}_{x}\pm i\mathbf{\Pi}_{y}$,
$\sigma_{\pm}=\sigma_{x}\pm i\sigma_{y}$ and $\mathbf{B}=\boldsymbol{\bigtriangledown}\times\mathbf{A}$.
The electrostatic potential $V(\mathbf{r})$ defines, \textit{e.g.},
the lateral confining potential of a 2D ballistic conductor which
defines the ring structure. One can obtain the one dimensional (1D)
Hamiltonian of a heavy hole in a ring following the procedure described
in Appendix, also outlined in Ref. \onlinecite{Meijer:jul2002}:
\begin{equation}
\begin{array}{l}
\hat{H}(\Phi)={\displaystyle \frac{\hbar\omega_{0}}{2}}\left(\widetilde{\partial}_{\varphi}\right)^{2}+{\displaystyle \frac{\hbar\omega_{B}}{2}}\sigma_{z}+\\
{\displaystyle \frac{\hbar\omega_{R}}{4}}\left(2(\cos3\varphi\sigma_{x}+\sin3\varphi\sigma_{y})\left(\widetilde{\partial}_{\varphi}\right)^{3}\right.\\
+i3(\cos3\varphi\sigma_{y}-\sin3\varphi\sigma_{x})\left[{\displaystyle \frac{3r_{0}^{2}}{w^{2}}}+1+3\left(\widetilde{\partial}_{\varphi}\right)^{2}\right]\\
+\left.(\cos3\varphi\sigma_{x}+\sin3\varphi\sigma_{y})\left[{\displaystyle \frac{6r_{0}^{2}}{w^{2}}}-7\right]\widetilde{\partial}_{\varphi}\right)\end{array}\label{ham1}\end{equation}
 where $\widetilde{\partial}_{\varphi}=\left(i\frac{\partial}{\partial\varphi}+\Phi\right)$,
$r_{0}$ is the radius of the ring, $w$ is the half width of the
ring channel, $\omega_{0}={\displaystyle \frac{\hbar}{mr_{0}^{2}}}$,
$\omega_{B}={\displaystyle \frac{g\mu_{B}B_{z}}{\hbar}}$, $\omega_{R}={\displaystyle \frac{2\alpha}{\hbar r_{0}^{3}}}$,
$\Phi={\displaystyle \frac{\pi r_{0}^{2}B}{hc/e}}$. Here we follow
the notation of Ref. \onlinecite{Frustaglia:jun2004} for easier
comparison. Note that the Hamiltonian Eq. (\ref{ham1}) is Hermitian
since the original Hamiltonian used in Appendix is Hermitian.

The general form of an eigen state of the Hamiltonian Eq. (\ref{ham1})
reads:\[
\Psi_{n}(\varphi)=e^{in\varphi}\left(\begin{array}{c}
\chi_{1}\\
{\displaystyle {\chi_{2}e}^{i3\varphi}}\end{array}\right),\]
 where the constants $\chi_{1(2)}$ do not depend on the angle $\varphi$.
By diagonalizing the corresponding matrix equation for $\chi_{1(2)}$,
we can obtain the eigenenergies and eigenstates.

The complete expressions for the eigenstates and their eigenenergies
are too cumbersome to be reproduced here, we thus present analytical
results for the two most important limits; (i) thin channel rings
with $k_{F}w<1$, and (ii) thick channel rings, $k_{F}w>1$,
with small Fermi length compared to the radius, $k_{F}r_{0}\gg1$
(this limit is usually realized in experiments).\cite{Konig:feb2006,Habib:2006}
In case (i) of a ring with a very thin channel, the Hamiltonian simplifies
to:\begin{equation}
\begin{array}{l}
\hat{H}(\Phi)={\displaystyle \frac{\hbar\omega_{0}}{2}}\left(\widetilde{\partial}_{\varphi}\right)^{2}+{\displaystyle \frac{\hbar\omega_{B}}{2}}\sigma_{z}+{\displaystyle \frac{\hbar\omega_{R}}{2}{\displaystyle \frac{3r_{0}^{2}}{w^{2}}}}\\
\times\left((\cos3\varphi\sigma_{x}+\sin3\varphi\sigma_{y})\widetilde{\partial}_{\varphi}+{\displaystyle \frac{3i}{2}}(\cos3\varphi\sigma_{y}-\sin3\varphi\sigma_{x})\right)\end{array},\label{ham2}\end{equation}
 with (non-normalized) eigenstates and eigenenergies:\begin{equation}
\Psi_{n}(\varphi)=e^{in\varphi}\left(\begin{array}{c}
{\displaystyle \frac{3+2n-2\Phi+{\displaystyle \frac{2}{3}}h\pm\lambda}{Q_{t}(3+2n-2\Phi)}}\\
{\displaystyle e^{i3\varphi}}\end{array}\right),\label{EigenSt1}\end{equation}
 \begin{equation}
E_{n}={\displaystyle \frac{\hbar\omega_{0}}{4}}\left((3+2n-2\Phi)^{2}/2+{\displaystyle \frac{9}{2}}\pm3\lambda\right),\label{root1}\end{equation}
where \[
\lambda=\sqrt{{\displaystyle \frac{4}{9}}h^{2}+{\displaystyle \frac{8}{3}}h(\Phi-n-{\displaystyle \frac{3}{2}})+(1+Q_{t}^{2})(3+2n-2\Phi)^{2}},\]
with $Q_{t}={\displaystyle \frac{r_{0}^{2}}{w^{2}}}{\displaystyle \frac{\omega_{R}}{\omega_{0}}}$
and $h=\omega_{B}/\omega_{0}$. We note however that this limit has
not been achieved yet experimentally, e.g. $wk_{F}\sim30$, although
perhaps an effectively narrower channel may be present in some experiments
due to irregularities in the ring.

Throughout this paper, we only consider the lowest transverse mode, which should be sufficient for answering the question of whether the larger frequency of conductance oscillations can be a result
of the cubic spin-orbit interaction. Thus, in the more experimentally relevant limit (ii), the largest terms in the
Hamiltonian Eq. (\ref{ham1}) can be captured by fixing the radial
coordinate in Eq. (\ref{cylinder}) to the average value $\left\langle R_{0}(r)\right| r \left|R_{0}(r)\right\rangle =r_{0}$, as it was done in Ref. \onlinecite{ARONOV:jan1993},
and consequently symmetrizing it (to make it Hermitian) by the following
procedure: \begin{equation}
\begin{array}{ll}
\hat{H}_{herm} & =(\hat{H}+\hat{H}^{\dagger})/2=\frac{\hbar\omega_{0}}{2}\left(\widetilde{\partial}_{\varphi}\right)^{2}+{\displaystyle \frac{\hbar\omega_{B}}{2}}\sigma_{z}\\
 & +{\displaystyle \frac{\hbar\omega_{R}}{4}}\left\{ (\cos3\varphi\sigma_{x}+\sin3\varphi\sigma_{y}),\left(\widetilde{\partial}_{\varphi}\right)^{3}\right\} \\
 & +i{\displaystyle \frac{3\hbar\omega_{R}}{2}}\left[(\cos3\varphi\sigma_{y}-\sin3\varphi\sigma_{x}),\left(\widetilde{\partial}_{\varphi}\right)^{2}\right]\\
 & +2\hbar\omega_{R}\left\{ (\cos3\varphi\sigma_{x}+\sin3\varphi\sigma_{y}),\left(\widetilde{\partial}_{\varphi}\right)\right\} \\
\\\end{array}\label{hermitian}\end{equation}
 where $\left\{ ...\right\} $ and $\left[...\right]$ mean commutator
and anticommutator respectively. Note that in the case of the Rashba
Hamiltonian considered in Ref. \onlinecite{Meijer:jul2002} there
is no difference between such symmetrization and the perturbative
procedure.

The (non-normalized) eigenstates of the Hamiltonian Eq. (\ref{hermitian})
are:\begin{equation}
\Psi_{n}(\varphi)=e^{in\varphi}\left(\begin{array}{c}
{\displaystyle \frac{\sqrt{12m+13}+{\displaystyle \frac{2}{3}}h\pm\lambda}{mQ\sqrt{12m+13}}}\\
{\displaystyle e^{i3\varphi}}\end{array}\right)\label{EigenSt2}\end{equation}
 where \[
\lambda=\sqrt{{\displaystyle \frac{4}{9}}h^{2}+{\displaystyle \frac{8}{3}}h(\Phi-n-{\displaystyle \frac{3}{2}})+(12m+13)(1+m^{2}Q^{2})},\]
 and $m=n^{2}/3+n-1/3+{\displaystyle \Phi(\Phi-2n-3)/3}$ and $Q=\omega_{R}/\omega_{0}$.
The associated eigenenergies read:\begin{equation}
E_{n}=\hbar\omega_{0}\left(6m+11\pm3\lambda\right)/4.\label{root2}\end{equation}
 Note that in general, the Hamiltonian Eq. (\ref{hermitian}) gives
six eigenstates for a fixed Fermi energy ($E_{F}=E_{n}$). In the
limit $\bigtriangleup\ll E_{F}$ (where $\bigtriangleup$ is the energy
of spin-orbit splitting and $E_{F}$ is the Fermi energy); however,
two of the six states have much larger number $n$, and correspond
to the unphysical situation when the cubic spin-orbit coupling term
dominates the spectrum creating an unphysical downturn in the spectrum,
which is truly not present. Hence, these two states are ignored on
the basis of this physical reason, i.e. they do not exists in the
physical system. It is convenient to describe the four conducting
states by increasing real numbers $n_{1-}\leq n_{2-}\leq n_{2+}\leq n_{1+}$,
solutions of the equation $E_{F}=E_{n}$ (see Eqs. (\ref{root1},\ref{root2})).

We consider a ring symmetrically coupled to two contact leads in order
to study the transport properties of the system subject to a low bias
voltage in the linear regime. To this end, we calculate the zero-temperature
conductance $G$ based on the Landauer formula: \begin{equation}
G=\frac{e^{2}}{h}\sum_{\nu',\nu=1}^{M}{\displaystyle T_{\nu'\nu}},\label{Landauer}\end{equation}
 where labels $\nu$ and $\nu'$ number the channel and spin. We assume
perfect coupling between leads and ring (\textit{i.e.}, fully transparent
contacts), neglecting backscattering effects leading to resonances.
In this approximation, the incoming spin $\left|\sigma\right\rangle $
propagates coherently along the four available channels, leaving the
ring in a mixed spin state $\left|\sigma_{out}\right\rangle =\sum_{i=1,2;s=\pm}\left\langle \Psi_{n_{is}}(0)\right.\left|\sigma\right\rangle \left|\Psi_{n_{is}}(\pi)\right\rangle $.
The spin-resolved transmission probabilities can be obtained by use
of a complete basis of incoming $\left|\sigma\right\rangle $ and
outgoing $\left|\sigma'\right\rangle $ spin states,\begin{equation}
G=\frac{e^{2}}{h}\sum_{\sigma'\sigma}|\left\langle \sigma'\right.\left|\sigma\right\rangle |^{2}.\label{spin}\end{equation}
 In sufficiently large rings, $k_{F}r_{0}\gg1$ (\textit{e.g.} $1/k_{F}\sim4\mbox{nm}$
in a HgTe QW with a heavy hole band)\cite{Zhang:jun2001}, the Zeeman
splitting for the magnetic fields considered is small compared to
other important energy scales. Summing over all spin-states in Eq.
(\ref{spin}) and disregarding the Zeeman term in Eq. (\ref{hermitian}),
we obtain the conductance: \begin{eqnarray}
G & = & \frac{e^{2}}{h}\left[1-\cos[\pi(n_{1+}-n_{2+})]\left(\frac{(A-1)}{2}\right.\right.\nonumber \\
 &  & +\left.\left.\frac{(A+1)}{2}\cos[2\pi\Phi]\right)\right]\label{conductance}\end{eqnarray}
 where $A=1$ in the limit (i) and $A={\displaystyle \frac{1+m_{1}m_{2}Q^{2}}{\sqrt{1+m_{1}^{2}Q^{2}}\sqrt{1+m_{2}^{2}Q^{2}}}}$
in the limit (ii), $(n_{1+}-n_{2+})$ is the difference between two
roots of Eqs. (\ref{root1},\ref{root2}), and $m_{1(2)}=n_{1(2)}^{2}/3+n_{1(2)}-1/3+{\displaystyle \Phi(\Phi-2n_{1(2)}-3)/3}$.

\begin{figure}[t]
 \includegraphics[width=0.95\columnwidth]{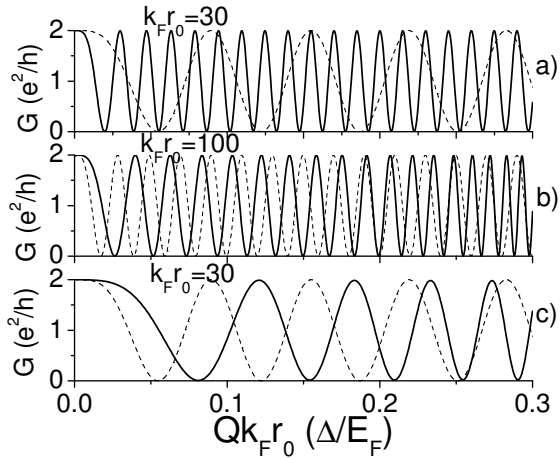}

\caption{Conductance-modulations in a 1D ring as a function of the dimensionless
spin-orbit strength $Qk_{F}r_{0}$ (note that state of the art experimental
systems are in a regime where $Qk_{F}r_{0}\lesssim0.1$); a) thin
heavy hole ring (solid line) is compared to the Rashba ring (dashed
line), $(k_{F}w)^{2}=1/2$; b) and c) thick heavy hole ring (solid
line) is compared to the Rashba ring (dashed line).}

\label{ring} 
\end{figure}

In the limit (i) of thin channel rings, we can find the difference
between the two roots:\[
n_{1+}-n_{2+}=3\sqrt{1+\left({\displaystyle \frac{r_{0}^{2}}{w^{2}}}Q\right)^{2}},\]
 which means that by making the ring channel thinner than the Fermi
length we can increase the frequency of conductance oscillations by
a factor of $3/(k_{F}w)^{2}$ (see Fig. (\ref{ring}a)). This results
from the increase in the spin-orbit splitting due to the quantization
in the radial direction. Experimental realization of thin channel
rings is very difficult and in the rest of the paper we concentrate
on the rings in the limit (ii) when $k_{F}w\gtrsim1$. Although such
rings should have more than one conducting channel, we suppose that
only one is important. This can be a result of the resonant transmission
of this channel, or incoherent transport through the other channels.

In the experimentally relevant limit (ii), for not too large $Q$,
we can approximate $(n_{1+}-n_{2+})$ in Eq. (\ref{conductance})
as:\begin{equation}
n_{1+}-n_{2+}\approx3+{\displaystyle \frac{2}{3}}\left(N_{F}(8+N_{F})-2\right)Q^{2}\label{argument}\end{equation}
 where $N_{F}=E_{F}/(\hbar\omega_{0})=(k_{F}r_{0})^{2}/2$.

\begin{figure}[t]
 \centerline{\includegraphics[width=0.55\columnwidth,angle=-90]{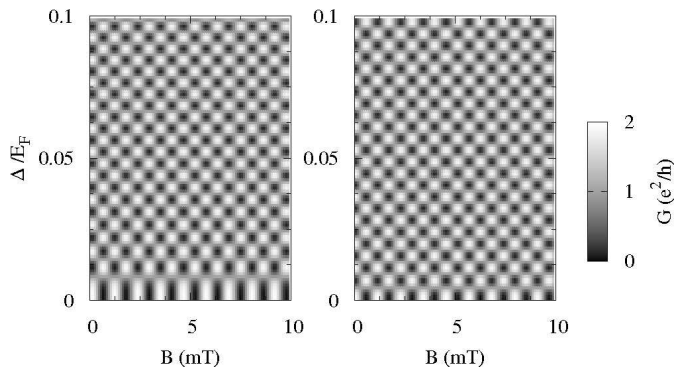}}

\caption{Conductance-modulations in an 1D ring as a function of magnetic field
and dimensionless spin-orbit strength (gate voltage). Left plot corresponds
to heavy hole spin-orbit interaction, right plot corresponds to Rashba
spin-orbit interaction; $r_{0}=1\mu\mbox{m}$, $1/k_{F}\sim4\mbox{nm}$,
$g=20$ and $m^{*}=0.031m$. Parameters of the left plot correspond
to the experimental setup in Ref. \onlinecite{Konig:feb2006}.}

\label{RingField} 
\end{figure}

\begin{figure}[t]
 \centerline{\includegraphics[width=0.55\columnwidth,angle=-90]{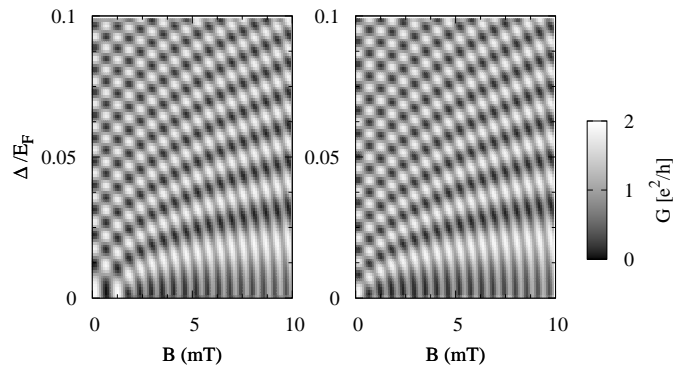}}

\caption{Conductance-modulations in an 1D ring as a function of magnetic field
and dimensionless spin-orbit strength (gate voltage) with enhanced
Zeeman splitting (multiplied by $10^{3}$). Left plot corresponds
to heavy hole spin-orbit interaction, right plot corresponds to Rashba
spin-orbit interaction; $r_{0}=1\mu\mbox{m}$, $1/k_{F}\sim4\mbox{nm}$,
$g=20$ and $m^{*}=0.031m$. Parameters of the left plot could correspond
to the experimental setup in Ref. \onlinecite{Konig:feb2006} with
the magnetic field (range $1-10\:\mbox{T}$) applied under some small
angle.}

\label{RingFieldZ} 
\end{figure}

When the parameter $A=1$, Eq. (\ref{conductance}) is also valid
for an electron ring considered in Ref. \onlinecite{Frustaglia:jun2004}.
This can be obtained by using Eq. (\ref{spin}) and electron Hamiltonian
considered in Ref. \onlinecite{Frustaglia:jun2004}. For the Rashba
ring, the difference between roots can be calculated exactly $n_{1+}-n_{2+}=\sqrt{1+Q_{e}^{2}}$
and $Q_{e}={\displaystyle \frac{2m\alpha_{e}r_{0}}{\hbar^{2}}}$ ($\alpha_{e}$
is the Rashba coupling parameter that differs from the one used in
Eq. (\ref{ham})). When the spin-orbit splittings in the hole ($\Delta/E_{F}\sim\sqrt{2N_{F}}Q$)
and electron ($\Delta_{e}/E_{F}\sim Q_{e}/\sqrt{2N_{F}}$) systems
match, we can write $Q_{e}\sim2N_{F}Q$. Therefore, the conductance
oscillations as a function of spin-orbit splitting for the electron
and hole systems have comparable periods (see Figs. \ref{ring} and
\ref{RingField})). The period of the hole system has a tendency to
become shorter as the spin-orbit splitting becomes larger (see Fig.
\ref{ring}) which is not the case for Rashba rings. Notably, a hole
(electron) does not develop sufficient phase difference in the case
when the ring radius is small compared to the Fermi length, as it
can be seen from Fig. \ref{ring}. However, as pointed out before,
in realistic systems we always have $k_{F}r_{0}\gg1$ ($1/k_{F}\sim4\mbox{nm}$
for a HgTe QW with a heavy hole band).\cite{Zhang:jun2001} In Fig.
\ref{RingField}, we plot the conductance oscillations in the HH ring
(left plot) compared to the Rashba ring (right plot) as a function
of the external magnetic field. Here the small g-factor of the electron
system is assumed to be the same as for the hole system for easier
comparison. Where as in the field direction there is not a large difference
in the conductance fluctuations, the changing oscillation frequency
of the hole system becomes more obvious as compared to the electron
system.

We next take into account the Zeeman splitting as a first order correction.
The perturbed eigenenergies become: \begin{eqnarray}
E_{n} & = & \frac{\hbar\omega_{0}}{4}\left(6m+11\pm3\sqrt{(12m+13)(1+m^{2}Q^{2})}\right)\nonumber \\
 &  & \mp{\displaystyle \frac{h\:\mbox{sign}(n)}{2\sqrt{1+m^{2}Q^{2}}}}.\label{root3}\end{eqnarray}
 where $h=\omega_{B}/\omega_{0}$, $\mbox{sign}$ is the sign function.
To the first order in the spin-orbit interaction and Zeeman splitting,
Eq. (\ref{conductance}) can still describe the conductance after
the following substitution: \begin{eqnarray}
n_{1+}-n_{2+} & \rightarrow & n_{1+}-n_{2+}+\frac{h}{\sqrt{1+\bar{m}^{2}Q^{2}}}/\left(\frac{\partial E_{n}}{\partial n}\right)\nonumber \\
 & \approx & n_{1+}-n_{2+}-\frac{h}{\bar{n}\sqrt{1+\bar{m}^{2}Q^{2}}},\label{conductance1}\end{eqnarray}
 where $\bar{n}$ is the average of $n_{1+}$ and $n_{2+}$ and $\bar{m}(\bar{n})$
is defined the same way as in Eq. (\ref{EigenSt2}). For small Zeeman
splittings ($h/n\ll1$), which holds for realistic rings, the conductance
is well described by Eq. (\ref{conductance}) and the chessboard pattern
in Fig. \ref{ring}.

We present the results of calculations for larger Zeeman splittings
($h/n\sim1$) in Fig. \ref{RingFieldZ}. The analytical expressions
are too cumbersome and we do not reproduce them here. As one can see,
the Zeeman term can substantially delay the development of Aharonov-Casher
oscillations, especially for larger magnetic fields. In order to experimentally realize
this situation, one may apply much larger magnetic
fields at some angle to the plane of the ring. Such procedure diminishes
the magnetic flux through the structure, allowing to work at higher
magnetic fields with much larger Zeeman splittings.

Given the fact that for the experiments in Ref. \onlinecite{Konig:feb2006}
and \onlinecite{Habib:2006} the experimental systems are in a regime
where $Qk_{F}r_{0}\lesssim0.1$ and $k_{F}w>1$, the frequency of
conductance oscillation expected for a single mode (1D) ring are of
similar order for both the hole and electron systems. %NOTE: add comment on Hartmun
We conclude that the multichannel analysis of the experiments is an
important feature for understanding them at present. The increasing
frequency of oscillation observed in our calculation, only seen theoretically
in the hole gas systems, will require a strength of doping and confining
electric field which has not been experimentally achieved at present.

\textit{Acknowledgments}. The authors are grateful for useful discussions
with B. Habib, and M. Shayegan. This work was supported by ONR under
Grant No. ONR-N000140610122 and by the NSF under Grant no. DMR-0547875,
by the EPSRC through Grant No. GR/S81407/01, by the Grant Agency and
Academy of Sciences of the Czech Republic through Grants No. 202/05/0575,
and AV0Z10100521, by the Ministry of Education of the Czech Republic
Grant No. LC510, by the EU Project NANOSPIN FP6-2002-IST-015728, and
from the EU EUROCORES Project SPICO FON/06/E002. M.F.B. is supported
by the Department of Education through a GAANN fellowship. Jairo Sinova
is a Cottrell Scholar of Research Foundation.

\appendix

\section{Derivation of the 1D Hamiltonian}

In this Appendix, we present the derivation of the 1D Hamiltonian
for the hole ring. In cylindrical coordinates, with $x=r\cos\phi$
and $y=r\sin\phi$, Eq. (\ref{ham}) reads \begin{widetext} \begin{equation}
\begin{array}{l}
\widehat{H}(r,\phi)=-\frac{\hbar}{2m}\left(\frac{\partial^{2}}{\partial r^{2}}+\frac{\partial}{r\partial r}-\frac{1}{r^{2}}\left(i\frac{\partial}{\partial\phi}+\Phi\right)^{2}\right)+V(r)+\frac{\alpha i}{\hbar r^{3}}\text{cos}[3\phi]\left(3r\left(\sigma_{y}-i\sigma_{x}\Phi+\sigma_{y}\Phi^{2}\right)\frac{\partial}{\partial r}-3r^{2}(\sigma_{y}-i\sigma_{x}\Phi)\frac{\partial^{2}}{\partial r^{2}}\right.\\
+r^{3}\sigma_{y}\frac{\partial^{3}}{\partial r^{3}}-8\sigma_{x}\frac{\partial}{\partial\phi}-6i\sigma_{y}\Phi\frac{\partial}{\partial\phi}+9r\sigma_{x}\frac{\partial^{2}}{\partial\phi\partial r}+6ir\sigma_{y}\Phi\frac{\partial^{2}}{\partial\phi\partial r}-3r^{2}\sigma_{x}\frac{\partial^{3}}{\partial\phi\partial r^{2}}+6\sigma_{y}\frac{\partial^{2}}{\partial\phi^{2}}\left.-3r\sigma_{y}\frac{\partial^{3}}{\partial\phi^{2}\partial r}+i\sigma_{x}\left(i\frac{\partial}{\partial\phi}+\Phi\right)^{3}\right)\\
-\frac{\alpha i}{\hbar r^{3}}\text{sin}[3\phi]\left(3r\left(\sigma_{x}+i\sigma_{y}\Phi+\sigma_{x}\Phi^{2}\right)\frac{\partial}{\partial r}-3r^{2}(\sigma_{x}+i\sigma_{y}\Phi)\frac{\partial^{2}}{\partial r^{2}}\right.+r^{3}\sigma_{x}\frac{\partial^{3}}{\partial r^{3}}+8\sigma_{y}\frac{\partial}{\partial\phi}-6i\sigma_{x}\Phi\frac{\partial}{\partial\phi}-9r\sigma_{y}\frac{\partial^{2}}{\partial\phi\partial r}+6ir\sigma_{x}\Phi\frac{\partial^{2}}{\partial\phi\partial r}\\
+3r^{2}\sigma_{y}\frac{\partial^{3}}{\partial\phi\partial r^{2}}+6\sigma_{x}\frac{\partial^{2}}{\partial\phi^{2}}\left.-3r\sigma_{x}\frac{\partial^{3}}{\partial\phi^{2}\partial r}-i\sigma_{y}\left(i\frac{\partial}{\partial\phi}+\Phi\right)^{3}\right)\end{array}\label{cylinder}\end{equation}
 \end{widetext} where $\Phi$ is the magnetic flux through the ring
as a function of the radial coordinate, $\Phi={\displaystyle \frac{\pi r^{2}B}{hc/e}}$.
We employ the perturbative method used in Ref. \onlinecite{Meijer:jul2002}
by separating the Hamiltonian Eq. (\ref{cylinder}) into the dominant
part:\[
\widehat{H}_{0}(r,\phi)=-\frac{\hbar}{2m}\left(\frac{\partial^{2}}{\partial r^{2}}+\frac{\partial}{r\partial r}-\frac{1}{r^{2}}\left(i\frac{\partial}{\partial\phi}+\Phi\right)^{2}\right)+V(r)\]
 and the remaining perturbation $\widehat{H}_{1}=\widehat{H}-\widehat{H}_{0}$.
In the limit $w\ll r_{0}$ the solution of the Hamiltonian $\widehat{H}_{0}$
can be found as a degenerate set of states $\Psi(r,\phi)=R_{0}(r)\Phi_{n}(\phi)$
where $R_{0}(r)$ is the lowest radial mode and $\Phi_{n}(\phi)$
is a spinor function of the angle $\phi$. It can be shown that the
degeneracy in spin space can be lifted by diagonalizing the following
Hamiltonian: \[
\widehat{H}_{1D}(\phi)=\left\langle R_{0}(r)\right|\widehat{H}_{1}+\frac{\hbar}{2mr^{2}}\left(i\frac{\partial}{\partial\phi}+\Phi\right)^{2}\left|R_{0}(r)\right\rangle ,\]
 which allows us to find the desired 1D Hamiltonian.

We use the lowest radial solution found in Ref. \onlinecite{Meijer:jul2002},
$R_{0}(r)={\displaystyle \left(\frac{1}{wr_{0}\sqrt{\pi/2}}\right)^{1/2}e^{-(r-r_{0})^{2}/w^{2}}}$,
leading to the following expectation values, $\left\langle R_{0}(r)\right|\frac{\partial}{\partial r}\left|R_{0}(r)\right\rangle =-1/(2r_{0})$,
$\left\langle R_{0}(r)\right|\frac{\partial}{r^{2}\partial r}\left|R_{0}(r)\right\rangle =1/(2r_{0})$,
$\left\langle R_{0}(r)\right|r^{2}\frac{\partial}{\partial r}\left|R_{0}(r)\right\rangle =-3r_{0}/2$,
$\left\langle R_{0}(r)\right|\frac{r\partial^{2}}{\partial r^{2}}\left|R_{0}(r)\right\rangle =1/(4r_{0})-r_{0}/w^{2}$,
$\left\langle R_{0}(r)\right|\frac{\partial^{2}}{r\partial r^{2}}\left|R_{0}(r)\right\rangle =-1/(r_{0}w^{2})$
and $\left\langle R_{0}(r)\right|\frac{\partial^{3}}{\partial r^{3}}\left|R_{0}(r)\right\rangle =3/(2r_{0}w^{2})$.
The Hermitian 1D Hamiltonian for the hole ring takes the form of Eq.
(\ref{ham1}).

\end{document}